\documentclass[12pt]{article}
\usepackage{graphicx}
\usepackage{cite}
\def\({\left(}
\def\){\right)}
\def\[{\left[}
\def\]{\right]}
\def\be{\begin{equation}}
\def\ee{\end{equation}}
\def\bea{\begin{eqnarray}}
\def\eea{\end{eqnarray}}
\def\ba{\begin{array}}
\def\ea{\end{array}}
\def\pa{\partial}

\def\nn{\nonumber}

\def\de{\delta}

\def\de{\delta}

\def\eps{\epsilon}

\def\vq{\vec{q}}
\def\vp{\vec{p}}

\def\nin{\noindent}

\begin{document}
\hfill\parbox{5cm}{\normalsize CERN-PH-TH/2004-024}

\unitlength1cm

\begin{center}

{\large {\bf Electronic contribution to the oscillations of a gravitational
antenna}}\\

\vspace*{0.3cm}
Vincenzo Branchina\footnote{Vincenzo.Branchina@cern.ch}

{\it Department of Physics, CERN, Theory Division, CH-1211  Geneva 23,  Switzerland}\\

and 

{\it IReS Theory Group, ULP and CNRS, 23 rue du Loess, 67037 Strasbourg, France}\\

\vspace*{0.2cm}
Alice Gasparini\footnote{Alice.Gasparini@physics.unige.ch}\\
{\it D\'epartement de Physique Th\'eorique, Universit\'e de Gen\`eve,\\
24 quai Ernest-Ansermet, CH-1211 Gen\`eve 4\\}

\vspace*{0.2cm}
Anna Rissone\footnote{Anna.Rissone@physics.unige.ch}\\
{\it D\'epartement de Physique Th\'eorique, Universit\'e de Gen\`eve,\\
24 quai Ernest-Ansermet, CH-1211 Gen\`eve 4\\}

\vspace*{0.7cm}

{\LARGE Abstract}\\

\end{center}

We carefully analyse the contribution to the oscillations of a metallic
gravitational antenna due to the interaction between the electrons of 
the bar and the incoming gravitational wave. To this end, we first 
derive the total microscopic Hamiltonian of the wave--antenna system 
and then compute the contribution to the attenuation factor
due to the electron-graviton interaction. As compared to the ordinary 
damping factor, which is due to the electron viscosity, this term 
turns out to be totally negligible. 
This result confirms that the only relevant mechanism for the interaction 
of a gravitational wave with a metallic antenna is its direct coupling 
with the bar normal modes.

\vskip 20pt

Pacs 4.30; 4.80; 95.55.Ym; 63.20.-e.

\newpage

\section{Introduction}\label{intro}

The detection of gravitational waves represents nowadays one of the most 
challenging and stimulating problems in experimental and theoretical 
physics.

At the beginning of the sixties, Weber pioneered a research program
based on the construction of metallic antennae (aluminium bars), and 
computed the cross-section for the absorption of gravitational 
waves\cite{web1} (see also \cite{ruf,wein}) under the assumption that 
the relevant mechanism 
that governs the transfer of energy is a direct resonant coupling 
between the wave and the bar normal modes, the so called resonant 
assumption\cite{wein}.  
The detector is so built that a signal is observed when the 
frequency of the incoming wave is close to the bar fundamental 
mode, the cross-section taking a Breit-Wigner shape
around this frequency. 

Weber successively proposed a correction to his own computation. 
Arguing that the bar has very many resonant frequencies within 
the tiny detection range, he found an enormous enhancement of 
the cross-section\cite{web2}. Later Preparata\cite{prep1} 
considered a different mechanism that again lead to an enhancement 
of the cross-section. The problems with both of these 
mechanisms have been clearly identified \cite{prep2,thor,gris}.  
It is by now largely accepted that, within the framework of the 
resonant assumption, the correct result is the old one \cite{web1}. 

However, in addition to the direct graviton-phonon coupling 
considered in \cite{web1, ruf, wein}, which at the microscopic level 
is due to a modification 
of the geodesic distance between the ions of the lattice in the 
presence of the gravitational wave, it is clear that the gravitons 
also couple to the electrons of the metal. Due to the electron-phonon 
interaction, an indirect coupling between gravitons and phonons is then
generated\footnote{It has been recently  claimed that this interaction 
is responsible for an enormous enhancement of the 
cross-section\cite{sri}. Our results (see section (\ref{atten})) 
strongly disagree with this conclusion.}.

By considering the 
microscopic Hamiltonian for the interaction of a gravitational wave with 
the bar, in this paper we carefully analyse the electronic contribution 
to the absorption of gravitational waves by the antenna. More specifically, 
we compute the contribution (renormalization) to the attenuation
factor, i.e. the imaginary part of the frequency of the propagating acoustic 
wave, due to the interaction of the electrons with the gravitational
wave. Depending on its sign, this term could lead to 
an attenuation or an amplification of the bar oscillation amplitude.
This is a correction to the ``ordinary" (positive) factor, called 
$\Gamma^0$ from now on, which is the bar attenuation factor in the 
absence of gravitational waves (see section \ref{atten} below).
Therefore, irrespectively of its sign, the actual 
relevance of such a term depends on its magnitude relative to
$\Gamma^0$. 

A similar phenomenon has already been considered for the
propagation of an acoustic wave in a semiconductor in the presence 
of an electromagnetic wave\cite{eps1,eps2}. 
Here the electron-phonon interaction generates, via electron-hole 
loops, an indirect photon-phonon coupling. Although this 
electron viscosity generally causes an attenuation of the 
acoustic wave, under certain circumstances the wave 
can be amplified. 
We shall see in the following that, while this  
is possible for the electromagnetic case, for a
gravitational wave hitting a metallic antenna such an amplification
cannot occur. 

Present bars are aluminium detectors operating at a temperature 
of about $2\, ^{\circ}\mathrm{K}.$ To avoid uncontrolled
complications due to superconductivity effects, the antennae 
operate at temperatures $T>T_{s}$, where $T_s$ is the 
transition temperature to the superconducting state. We also 
restrict ourselves to this case.

Let us focus now our attention on the graviton-phonon interaction,
which is the only interaction mechanism usually considered, and 
suppose that a gravitational burst hits perpendicularly the 
antenna. If the bar lies along, say, the $z$ axis, the wave excites 
the modes $\omega_{\vq}=\frac{\pi v_{s}}{L}(2n+1)$, 
where $\vq=\vq_{_n}=(0,0,\frac{\pi}{L}(2n+1)), $ $L$ is the length
of the bar, $v_{s}$ is the sound speed in aluminium, and $n$  
a positive integer. If, as we assume, the incoming wave $h_{ij}$ 
contains a superposition of frequencies having a cutoff which does 
not greatly exceed the kHz, the only excited mode is the 
fundamental one.

As is well known, the detector measures the energy stored 
in the bar, $E_s$, which is proportional to the 
square of the amplitude of the bar oscillations. This quantity 
is related
to the Fourier transform $\tilde{h}_{zz}$ of the signal
in the following way \cite{gwexp}  :

\be\label{es}
E_s=\frac{ML^2}{\pi^2}\omega^4_{\vq_{_0}}\vert\tilde{h}_{zz}
(\omega_{\vq_{_0}})\vert^2\, .
\ee

\nin
Here $\tilde{h}_{zz}$is the $zz$ component of the Fourier transform
of the gravitational wave $h_{ij}$, $M$ is the mass of the bar, 
$\omega_{\vq_{_0}}$ its fundamental frequency, and  $L$, as before, 
the length of the bar. Note that, for notational convenience, but 
without loss of generality, we have chosen the axes so that the antenna 
lies along the $z$ direction. Moreover we have considered a 
gravitational wave perpendicular to the direction of the antenna. 
This is why in Eq.(\ref{es}) only the $zz$ component of the 
Fourier transform of the gravitational wave appears. 

Eq.(\ref{es}) is obtained by considering the
graviton-phonon interaction only, and our problem 
amounts to the question of whether the indirect
graviton-phonon interaction induced by the electrons
can lead to a significant modification of this equation. 
In fact, would this additional interaction generate
a value of the attenuation factor significantly lower than 
$\Gamma^0$, the bar oscillation amplitude would turn out to be  
greater than expecteded (and $E_s$ higher). Therefore, 
taking into account only the direct interaction, we  
would be lead to an overestimate of the signal.

The oscillations induced in the bar by the gravitational wave
have very small  amplitudes ($\Delta L/L\sim 10^{-21}$), and
they have to be amplified by an electronic system. The noise
in transducing these bar oscillations into electronic signals
allows for the amplification only of those amplitudes that are
larger than a given threshold. Therefore, in the case of
resonance, the detector is practically blind to frequencies
other than those within a narrow band around the fundamental
one. In the following we consider an incoming plane wave with
frequency $\Omega\simeq \omega_{\vq_{_0}}$.

The rest of the paper is organised as follows.
In section \ref{elmw} we briefly sketch the derivation of the
attenuation factor for the electromagnetic case mentioned above. 
In section \ref{hamilto} we present the derivation of the 
microscopic Hamiltonian for the interaction of the gravitational 
wave with the antenna. In section \ref{gravw} the renormalization 
of $\omega_{\vq_{_0}}$ in the presence of the gravitational wave 
is discussed, and in section \ref{atten} we compute the attenuation 
factor. Section \ref{conclu} contains the summary and conclusions.

\section{Sound waves in the field of an
electromagnetic wave}\label{elmw}

To set up the tools for our investigation, and to introduce 
some notations, in this section we very briefly review the derivation 
of the correction to the attenuation factor for an acoustic wave that 
propagates in a 
semiconductor in the field of an electromagnetic wave, referring 
to\cite{eps1, eps2} for details.

The microscopic Hamiltonian of the system (external electromagnetic wave
+ electrons + phonons) is: 

\bea\label{ham}
H =\frac{1}{2 m}\sum_{\vec{p}}(\vec{p} - \frac{e}{c}\vec A (t))^2
a^{\dagger}_{\vec{p}} \, a_{\vec{p}} + \sum_{\vec{k}} \omega_{\vec{k}}
\,
b^{\dagger}_{\vec{k}} \, b_{\vec{k}} + \sum_{\vec{p}, \vec{k}}
C_{\vec{k}} \,
a^{\dagger}_{\vec{p} + \vec{k}} \, a_{\vec{p}}\, (b_{\vec{k}} +
b^{\dagger}_{-\vec{k}})
\eea

\nin 
where $a_{\vec{p}}$ and  $a^{\dagger}_{\vec{p}}$ are the annihilation and 
creation operators for the electrons, $b_{\vec{k}}$ and $b^{\dagger}_{\vec{k}}$
the annihilation and creation operators for the phonons, $m$ the electron 
mass, $C_{\vec{k}}$ the coupling constants of the electron-phonon
interaction, and $\vec A(t)$ is the vector potential of the incoming
electromagnetic wave: $\vec A(t) =\vec A_0\, {\rm cos}\,(\Omega t)$ 
(corresponding to $\vec E(t) =\vec E_0\, {\rm sin}\,(\Omega t)$). Note 
that we are considering an electromagnetic wave of wavelength 
large with respect to the linear dimension of the system, so that 
its spatial dependence, within the semiconductor itself, can be neglected. 

Following the standard procedure, we assume that in the infinite 
past, i.e. at ~$t=-\infty$,~ the electrons and the phonons do not interact 
and that the external field is absent.
If $\rho(t)$ is the density matrix of the system, and we denote 
quantum-statistical averages with standard notation (i.e. for a 
generic operator $\cal {O}$ we write 
${\rm Tr} [\rho(t) {\cal O}]={\langle {\cal O}\rangle}_{_t}$),  
the above condition for the phonons reads:
$\langle b_{\vec{k}}\rangle_{_{-\infty}} =0$ and 
$\langle b^{\dagger}_{\vec{k}}\rangle_{_{-\infty}} =0$.
The external electromagnetic field and the electron-phonon interaction
are then adiabatically switched on. At a certain time, with the help of 
an external source, phonons of given momentum, say $\vec q$, 
are excited. Starting from this time: 
$\langle b_{\vec{q}}\rangle_{t} \neq0$ and 
$\langle b^{\dagger}_{\vec{q}}\rangle_{t} \neq0$. 

The renormalization of the phonon attenuation factor, 
$\Gamma_{\vec q}$, and more generally the renormalization 
of $\omega_{\vec q}$, can be obtained by considering the time 
evolution equation for 
$\langle b_{\vec{q}}\rangle_{t}$ \footnote{Alternatively, we could 
compute the renormalised frequency by considering 
the Feynman diagrams for the phonon propagator. Note that, due 
to the presence of the external electromagnetic field, see the first
term in Eq.(\ref{ham}), there are additional loop corrections 
containing insertion of external electromagnetic lines. Eq.(\ref{sig}) 
below gives an example of these additional terms. As it is immediately
clear from its form, this is the contribution to the renormalization 
of $\omega_{\vec q}$ where the fermion loop contains the insertion of 
an external photon line (absorption and emission  
of a photon with energy $\hbar\Omega$).}. 
This equation is easily obtained  
with the help of the Liouville equation for the density matrix:

\be\label{liouv}
i\partial_{t}\rho=[H,\rho]. 
\ee

For the Hamiltonian (\ref{ham}), we have from Eq.(\ref{liouv}):

\be\label{eqb}
i\partial_{t}\langle b_{\vec{q}}\rangle_{t}
-\omega_{\vec{q}}\,\langle b_{\vec{q}}\rangle_{t} =C_{\vec{q}}\,
\sum_{\vec p}\langle a^{\dagger}_{\vec{p}-\vec{q}}
\,a_{\vec{p}}\rangle_{t\, .}
\ee

\nin
The r.h.s of Eq.(\ref{eqb}) provides a small correction, due to 
the electron-phonon interaction, to the free equation: 
$i\partial_{t}\langle b_{\vec{q}}\rangle_{t}
-\omega_{\vec{q}}\,\langle b_{\vec{q}}\rangle_{t}=0$, whose elementary
solution is: 
$\langle b_{\vec{q}}\rangle_{t}= {\rm A}\, e^{-i\omega_{\vec q} t}$, with
${\rm A}$ an integration constant.

Going to Fourier space, i.e. writing 
$\langle b_{\vec{q}}\rangle_{t}=\int d \omega B(\omega) e^{-i\omega t}$, 
we derive from Eq.(\ref{eqb}) an equation for $B(\omega)$. Finally, 
by considering a value of $\omega$ close to $\omega_{_{\vq}}$, it is not 
difficult to see that from this last equation we 
can obtain the value of the phonon frequency, $\bar \omega_{\vec q}$, 
renormalised by the electron-phonon 
interaction\cite{eps2}. Moreover, by 
simple inspection of the free solution, we see that the attenuation 
factor for the acoustic wave is given by the negative of 
the imaginary part of $\bar \omega_{\vec q}$. 

In order to extract $\bar \omega_{\vec q}$ from Eq.(\ref{eqb}), 
a certain number of steps are needed (see \cite{eps2} for the details). 
First, again with the help of the Liouville equation, an evolution 
equation for the correlator $\langle a^{\dagger}_{\vec{p}-\vec{q}} 
\,a_{\vec{p}}\rangle_{t}$, which appears in the r.h.s. of Eq.(\ref{eqb}),
is derived. As the r.h.s. of Eq.(\ref{eqb}) already
contains one power of the small electron-phonon coupling, we can limit
ourselves to consider the lowest order solution in $C_{\vec q}$ to 
this new equation. This approximated solution contains the factor 
$\langle b_{\vec{q}}\rangle_{t}$. Therefore, from 
Eq.(\ref{eqb}), going again to Fourier space and dropping the factor 
$B(\omega)$, we get (to the lowest order in $C_{\vec q}$) \cite{eps2}:

\be\label{omegaa}
\bar \omega_{\vec q} =\omega_{\vec q} + C^2_{\vec q}\, 
\Sigma(\omega_{\vec q})\, ,
\ee

\nin 
where $\Sigma(\omega_{\vec q})$ contains the ``ordinary"
contribution to the renormalization of $\omega_{\vec q}$, 
$\Sigma^0$, as well as the additional term due 
to the presence of the electromagnetic wave ($\Sigma^{el}$) : 
 
\be\label{sigg}
\Sigma(\omega_{\vec q})=
\Sigma^0(\omega_{\vec q}) + 
\Sigma^{el}(\omega_{\vec q})\, .
\ee

For our purposes, we are mainly interested in 
$\Sigma^{el}(\omega_{\vec q})$. To the lowest order 
in the dimensionless parameter 
$a=\frac{e \vec E_0\cdot \vec q}{m \Omega^2}$, it is \cite{eps2} :

\be\label{sig}
\Sigma^{el}(\omega_{\vec q})=\frac{a^2}{4}
\sum_{\vp}\(\frac{n_{\vp-\vq}-n_{\vp}}
{\eps_{\vp-\vq}-\eps_{\vp}+\hbar\bar{\omega}_{\vq}+\hbar\Omega+i\eps}+
\frac{n_{\vp-\vq}-n_{\vp}}
{\eps_{\vp-\vq}-\eps_{\vp}+\hbar\bar{\omega}_{\vq}-\hbar\Omega+i\eps}\)\,
,
\ee

\nin
where $n_{\vec{p}}$ the Fermi distribution function, 
$\eps_{\vec{p}}=\frac{\vec p^{\,2}}{2m}$, and $i \varepsilon$,
as usual, implements the appropriate boundary conditions.

From Eqs. (\ref{omegaa}) and (\ref{sig}), it is not difficult to see 
that, if $\hbar\Omega >> \frac{p_{F}^2}{2m}$ and $\hbar q>>p_{F}$
($p_{F}$ is the Fermi momentum),
the electromagnetic contribution to the attenuation factor (which 
is nothing but the
$\vec E_0$ dependent part of $- {\rm Im}\, \bar \omega_{\vec q}$), 
~$\gamma_{\vec q}$, vanishes when 
$p_{F} < m v_{s} - |\frac{\hbar q}{2}-\frac{m\Omega}{q}|$  
($v_{s}$ is the sound velocity in the semiconductor), while it is :

\be\label{gamma}
\gamma_{\vec q} = \frac{VC_{\vq}^2a^2}{4\pi\hbar^4}\frac{m^2 v_{s}}{\hbar q}
\Biggl(\frac{\hbar q}{2}-\frac{m \Omega}{q}\Biggr)\; s^{-1}
\ee
\nin
for $p_{_{F}} > m v_{s} + |\frac{\hbar q}{2}-\frac{m\Omega}{q}|$.

As is clear from Eq.(\ref{sigg}), the total attenuation 
factor is decomposed as: 
$\Gamma_{\vec q}=\Gamma^0_{\vec q}+\gamma_{\vec q}$, 
where $\Gamma^0_{\vec q}$ $(>0)$ is the ``ordinary" 
(i.e. in the absence of an electromagnetic wave) attenuation factor.  

From Eq.(\ref{gamma}) we see that, when the phonon frequency is small 
enough with respect to the frequency of the incoming photon, more 
precisely when $q \le \sqrt{\frac{2 m \Omega}{\hbar}}$, 
$\gamma_{\vec q}$~ becomes negative. 
Moreover, as it depends quadratically on $\vec E_0$, for sufficiently large 
values of the external field, it can give a significant contribution 
to $\Gamma_{\vec q}$. Under these conditions, an 
amplification of the acoustic wave is observed\cite{eps2, hust}. 

As we shall see in the following, in the case of a gravitational 
wave hitting a metallic bar these conditions are not met.

\section{Hamiltonian of the GW-antenna system}
\label{hamilto}

In this section we present the derivation of the microscopic 
Hamiltonian for a gravitational wave
interacting with a metallic antenna. 

The total energy-momentum of this system is:
${T}_{\mu\nu}={T}_{\mu\nu}^{\,\rm mat}+ 
{T}_{\mu\nu}^{\,\rm gw}$\, ,
where ${T}_{\mu\nu}^{\,\rm gw}$ is the pure gravitational contribution,
while ${T}_{\mu\nu}^{\,\rm mat}$ is the matter one, including
its interaction with the gravitational wave.
Being the metric just the flat background plus a small deviation, 
$g_{\mu\nu}=\eta_{\mu\nu}+h_{\mu\nu}$, we can consider the linearised 
expression for $T_{\mu\nu}^{\rm gw}$. Inserting this approximation 
in the energy-momentum 
conservation equation, $\pa^{\nu}T_{\mu \nu}=0$, we get  
(we use: $\eta_{\alpha\beta}={\rm diag}(+1,-1,-1,-1)$):

\be\label{cons}
\pa^{\nu}T_{\mu \nu}^{\rm mat}-\frac{1}{2}\pa_{\mu}h^{\alpha\beta}
T^{\rm mat}_{\alpha\beta}=0\, .
\ee

Clearly $T_{\mu \nu}^{\rm{mat}}$ contains 
the contributions of both the ions and the electrons: 

\be\label{enmom}
{T}_{\mu \nu}^{\rm mat} = {T}_{\mu \nu}^{\rm el}
+ {T}_{\mu \nu}^{\rm ion} + {T}_{\mu \nu}^{\rm el-ion},
\ee

\nin
where the last term in the r.h.s. is due to the electron-ion
interaction, while $T_{\mu \nu}^{\rm el}$ and $T_{\mu \nu}^{\rm ion}$ 
are the contributions from the non-interacting electron and ion systems
in the presence of the external gravitational field.  

Being the electron-ion interaction small, the Hamiltonian for the 
free electrons and ions can be derived by considering, at first, the 
electron gas and the ions as non-interacting. Moreover, once we keep  
only the the first two terms in the r.h.s. 
of Eq.(\ref{enmom}), we note that  
$T_{\mu \nu}^{\rm el}$ and $T_{\mu \nu}^{\rm ion}$, in this limit, 
separately satisfy Eq.(\ref{cons}).  

We could derive the complete Hamiltonian of the system from the 
linearised Eq.(\ref{cons}). In the following, however, we 
shall follow this pattern only to derive the Hamiltonian 
for the electrons 
interacting with the external gravitational field. 
In fact, by closely  parallelling \cite{prep2}, we shall  see that 
the Hamiltonian for 
the interaction of the phonons with the gravitational wave is easily
obtained from the expression for the 
force coming from the geodesic deviation equation\cite{gravitation}.
Finally, the free phonon and the electron-phonon 
interaction terms are the second 
and the third term in the r.h.s. of Eq.(\ref{ham}) respectively. 

As is clear from the above considerations, the total 
Hamiltonian of the system has the form:

\be\label{hamm}
{H} = {H}_{0}^{el} + {H}^{gw-el} + {H}_{0}^{ph} 
+ {H}^{gw-ph} + {H}^{el-ph}\, ,
\ee 

\nin
where ${H}_{0}^{el} + {H}^{gw-el}$ is the Hamiltonian for the
electrons in the presence of the gravitational wave, ${H}_{0}^{ph} 
+ {H}^{gw-ph}$ is the phonons' one, and ${H}^{el-ph}$ is the 
electron-phonon interaction term.   

Let us consider now Eq.(\ref{cons}) for $T_{\mu \nu}^{\rm el}$.
As we are neglecting the mutual interaction between the electrons,
it is sufficient to consider a single electron in the field
of the gravitational wave. The Hamiltonian of the electron 
non-interacting gas is simply the sum of the individual terms. 

Taking the spatial integral of Eq. (\ref{cons})
for $T_{\mu \nu}^{\rm el}$, we obtain
(dropping the superscript ``${\rm el}$" for simplicity):

\be
\frac{d}{dt}p_\mu =\frac{1}{2}
\int d^3 x\pa_{\mu}h_{\alpha\beta}T^{\alpha\beta}\, ,
\label{cons1}
\ee

\nin
where a vanishing total spatial divergence term has been omitted, and
$\int d^3 x T^0_{\mu}$ has been written as $p_\mu$. This is
clearly the equation of motion for the electron in the external
gravitational field. Our goal here is to find the Lagrangian, 
and then the Hamiltonian, from which this equation can be 
derived (see for instance \cite{ruf2}). 

As we are considering the linearised theory, the electron 
energy-momentum tensor can contain only the free term plus, 
possibly, an additional $O(h_{\mu\nu})$ term. We have then:

\be\label{tenso}
T^{\alpha \beta}= m u^{\alpha} u^{\beta}\de^3(\vec{x}-\vec{x}(t)) 
+ O(h_{\mu\nu}),
\ee

\nin where $m$ is the electron mass, $u^{\alpha}$ the electron 
four-velocity,  and $\vec{x}(t)$ the electron trajectory.

Inserting Eq.(\ref{tenso}) in  Eq.(\ref{cons1}), and keeping 
terms up to $O(h_{\mu\nu})$, we have:

\be\label{mov}
\frac{d}{dt}p_\mu =\frac{m}{2}u^{\alpha} u^{\beta}\pa_{\mu} h_{\alpha\beta}.
\ee

It is now a trivial exercise to show that the above equation of motion
can be obtained from the Lagrangian: 

\be\label{lag2}
L=\frac{m}{2}\(\eta_{\alpha\beta}+h_{\alpha\beta}\)u^{\alpha}u^{\beta}
\ee

\nin
and that  
\be\label{mom}
p_{\mu}=\frac{\pa L}{\pa u^{\mu}}=m\(h_{\alpha\mu}u^{\alpha}
+\eta_{\alpha\mu}u^{\alpha}\).
\ee

Choosing the Transverse Traceless (TT) gauge for $h_{\mu\nu}$, and remembering that,
as the electron gas is non-relativistic, the spatial components 
of the electron four-velocity, $u^{i}$, are nothing but the 
components of electron velocity, $v^i$, while $u^{0}\cong 1$, the above 
Lagrangian (up to an irrelevant constant term) becomes:

\be\label{lag3}
L=\frac{m}{2}\(\de_{ij}+h_{ij}\)v^{i}v^{j}\, .
\ee

\nin
The corresponding Hamiltonian is: 

\be\label{ham2}
H=\frac{\pa L}{\pa u^{i}}u^{i}-L=
\frac{1}{2m}\(\vec{p}^{\,\,2} - h_{ij}p^ip^j\),
\ee

\nin
and $p^i$ is the momentum canonically conjugate to the electron position.
Note that, as the wavelength of the typical incoming gravitational wave is 
large with respect to the linear dimension of the bar, $h_{ij}$ can be 
considered as spatial independent. 

The classical Hamiltonian of the non-relativistic, non-interacting, 
electron gas in 
the field of the incoming gravitational wave is the sum of 
terms of the form (\ref{ham2}), and the corresponding quantum
Hamiltonian is:

\be\label{ham3}
{H}_{0}^{el} + {H}^{gw-el} =  
\frac{1}{2 m}\sum_{\vec{p}}(\vec{p}^{\,\,2} - h_{i j} p^i p^j)
a^{\dagger}_{\vec{p}} a_{\vec{p}} \, .
\ee

Let us now turn our attention 
to ${H}^{gw-ph}$. As we said before, rather than following 
the same line of reasoning that lead to Eq.(\ref{ham3}), we 
derive this 
interaction term from the geodesic deviation expression for 
the force acting upon a generic point of mass $m$ \cite{gravitation}:
\be\label{force}
F_j = -m R_{j 0 k 0} x_k,
\ee 
where $x_k$ are the coordinates of the point and 
$R_{\mu \nu \rho \sigma}$ is the Riemann tensor of the metric 
field. In the TT gauge, again within the linear approximation, 
it is:
\be\label{riem}
 R_{j 0 k 0} = - \frac{1}{2} \ddot{h}_{j k}\, ,
\ee
while all the other components vanish.

Obviously the Hamiltonian corresponding to 
the force (\ref{force}) is \cite{prep2}:

\be\label{hamint1p}
H = \frac{m}{2} R_{j 0 k 0} x_j x_k \, .  
\ee

Now we choose the coordinate system  so that the bar lies along the 
$z$ axis and the origin coincides with the center of the bar.
Moreover, in order to face the most favourable conditions for the detection, 
we assume that the incoming wave propagates 
perpendicularly to the $z$ axis, say along the $x$ axis. Under 
these conditions, the only non vanishing components of $h_{ij}$ are:

\be\label{compo}
h_{yy}=-h_{zz}=-h ~~~~~~ {\rm and} ~~~~~~ h_{yz}=h_{zy}.~~~~~~~ 
\ee

For the purposes of our 
present analysis, we can neglect the $xy$ (circular) section of the 
bar, so that we can consider it as being essentially a unidimensional
chain of coupled harmonic oscillators (the ions). 
From Eqs.(\ref{riem}) and (\ref{hamint1p}), the 
interaction Hamiltonian between the ions and the gravitational wave can
be written as:

\be\label{iongra}
H^{\rm{ion-gw}} = - \frac{m_{\rm{ion}}}{4} \ddot{h} \sum_n (z_n)^2
\ee
where the sum is over the ion sites and $z_n$ is the position of the $n$-th
ion, which is given by:
\be
z_n = n a + \xi_n \, ,
\ee
where $a$ is the lattice spacing, and $ \xi_n$ is the displacement from
the equilibrium position of the $n$-th ion. To first order in the 
displacements, we have:
\be\label{h-ion-grav}
H^{\rm{ion-gw}} = - \frac{m_{\rm{ion}}}{2} a \ddot{h} \sum_n n \xi_n \, .
\ee

This Hamiltonian can be immediately written in terms of phonons if, as
usual, we develop $\xi_n$ in normal modes:
\be\label{norm-modes}
\xi_n = \frac{1}{\sqrt{N}} \sum_k \tilde{\xi}_k e^{i k a n}\, , 
\ee
where $N \gg 1$ is the number of ions, and the operators
$\tilde{\xi}_k$ satisfy the relations $\tilde{\xi}^{\dagger}_k =
\tilde{\xi}_{-k}$. By considering periodic boundary conditions, 
we have $k = \frac{2 \pi n_k}{N a}$,
where $n_k$ is an integer. 
 
The Hamiltonian for the interaction between the gravitational wave and
the phonons is now found by replacing Eq.(\ref{norm-modes}) into
Eq.(\ref{h-ion-grav}) and writing $\tilde{\xi}_k$ in terms of
creation and annihilation phonon operators \cite{book}:
\be
\tilde{\xi}_k = \sqrt{\frac{\hbar}{2 m_{ion} \omega_k}}\(b_k +
b^{\dagger}_{-k}\)\, . 
\ee
Performing the sum over $n$, in the large $N$ limit we have:

\be\label{somma}
\sum_{n} n e^{i k a n} = -i N^2 \frac{(-1)^{n_k}}{2 \pi n_k}\, .
\ee

\nin
Inserting now Eq.(\ref{somma}) in Eq.(\ref{h-ion-grav}), and replacing 
$N a$ with $L$, the length of the bar,  and $N m_{ion}$ with $M$, 
the mass of the bar, we get:

\be\label{ham4}
H^{\rm{gw-ph}} = - \ddot{h}(t) \sum_k
\frac{\alpha_k}{\sqrt{\omega_k}} \(b_k + b^{\dagger}_{-k}\) \, ,
\ee
with
\be
\alpha_k = -i \frac{(-1)^{n_k} L}{4 \pi n_k} \sqrt{\frac{\hbar M}{2}} \, . 
\ee

Finally, inserting Eqs.(\ref{ham3}) and (\ref{ham4}) in Eq. (\ref{hamm}),
and remembering that the terms ${H}_{0}^{ph}$ and  ${H}^{el-ph}$ of 
this equation are the second and the third term in the r.h.s. of 
Eq. (\ref{ham}) respectively, we can now write the total microscopic 
Hamiltonian for the interaction between the gravitational wave and 
the antenna. It is:

\bea\label{ham-tot}
H & = & \frac{1}{2 m}\sum_{\vec{p}}(\vec{p}^{\, 2} - h_{i j}(t) p^i p^j)
a^{\dagger}_{\vec{p}} a_{\vec{p}} + \sum_{\vec{k}} \omega_{\vec{k}}
b^{\dagger}_{\vec{k}}b_{\vec{k}} ~~~~~~~~~~~~~~~~~ \nn\\
& & +\sum_{\vec{p}, \vec{k}} C_{\vec{k}}
a^{\dagger}_{\vec{p} + \vec{k}} a_{\vec{p}}\, (b_{\vec{k}} +
b^{\dagger}_{-\vec{k}}) - \ddot{h}(t) \sum_{\vec k}
\frac{\alpha_{k}}{\sqrt{\omega_{k}}} (b_{\vec k} +
b^{\dagger}_{-\vec k})\, ,
\eea

\nin
where $h_{i j}$ and $h$ are given in Eq.(\ref{compo}). 

Having at our disposal the microscopic Hamiltonian of the system, 
we can now move to the central issue of the present work, namely the
computation of the contribution to the attenuation factor due to 
the interaction between the incoming gravitational wave and the 
electrons of the metallic antenna. To this end, 
following the pattern illustrated in section \ref{elmw}, we shall
first
consider the time evolution of ${\langle b_{{\vec q}_{_0}}\rangle}_t$,
where ${\vec q}_{_0}$ is the wave number of the fundamental mode, 
and then compute the corresponding attenuation factor. This is the 
subject of the two following sections.

\section{Time evolution of ${\langle b_{\vec q_{_0}}\rangle}_t$ and
renormalization of $\omega_{\vq_{_{0}}}$   } \label{gravw}

Let us consider a gravitational wave, with frequency $\Omega$ of the 
order of the $KHz$, hitting the metallic antenna. The wavelength of 
such a wave is much larger than the linear dimension of the bar, 
which is typically of about one meter. Therefore, the spatial 
dependence of $h_{ij}$ can be neglected and we can write:

\bea
h_{ij}=A_{ij}\cos(\Omega t)\label{hij}\, .
\eea

As usual (see section \ref{elmw}), we assume that in the infinite past
the gravitational wave is absent and that the electrons and the 
phonons do not interact. Therefore, for each value of $\vec k$,
$\langle b_{\vec{k}}\rangle_{-\infty} =0$.
The evolution equation for $\langle b_{\vq}\rangle_{t}$ 
(for notational simplicity, from now on we write ${\vec q}$
rather than ${\vec q}_{_0}$ to indicate the fundamental mode) 
is obtained with the help of the Liouville equation
(Eq.(\ref{liouv})). With the Hamiltonian (\ref{ham-tot}) 
we obtain:

\be\label{eomb}
i\partial_{t}\langle b_{\vq}\rangle_{t}
-\omega_{\vq}\,\langle b_{\vq}\rangle_{t} = \frac{\alpha_{-\vq}\, \Omega^2
    h(t)}{\sqrt{\omega_{\vq}}} + C_{\vq}\,
\sum_{\vec p}\langle a^{\dagger}_{\vec{p}-\vq}
\,a_{\vec{p}}\rangle_{t}  \, .
\ee

The correlator
$\langle a^{\dagger}_{\vec{p}-\vec{q}}\,a_{\vec{p}}\rangle_{t}$
in the r.h.s. of Eq.(\ref{eomb}) comes from the electron-phonon
interaction term. Again with the help of Eq.(\ref{liouv}), an 
equation for this correlator can be derived:

\bea\label{eoma}
i\partial_{t}\langle a^{\dagger}_{\vec{p}-\vec{q}}a_{\vec{p}}\rangle_{t} &-&
(\epsilon_{\vec{p}}-\epsilon_{\vec{p}-\vec{q}}-\frac{h_{ij}}
{2m}(2p_{i}q_{j}-q_{i}q_{j}))
\langle a^{\dagger}_{\vec{p}-\vec{q}}a_{\vec{p}}\rangle_{t}\nonumber \\
&=&\sum_{k}C_{k}\langle (a^{\dagger}_{\vec{p}-\vec{q}}a_{\vec{p}-\vec{k}}
-a^{\dagger}_{\vec{p}-\vec{q}+\vec{k}}a_{\vec{p}})
(b_{\vec{k}}+b^{\dagger}_{-\vec{k}})\rangle_{t}\, .
\eea

\nin
As it was expected, the right hand side of this equation
contains correlators of products of three operators, as for instance
the term 
$\langle a^{\dagger}_{\vec{p}-\vec{q}}a_{\vec{p}-\vec{k}}
b_{\vec{k}}\rangle$, and this is due to the electron-phonon coupling. 
In fact, repeatedly exploiting 
the Liouville equation, an infinite system of coupled differential 
equations for the different correlators is generated. Therefore, 
we have to resort to a suitable 
truncation of this system. In the following we shall see how such 
an approximation can be obtained.

The hypothesis that in the infinite past the electron gas is
noninteracting yields the boundary condition:
$\langle a^{\dagger}_{\vec{p}-\vec{q}} \,a_{\vec{p}}\rangle_{-\infty}=0$.
Solving (formally) Eq.(\ref{eoma}) under this condition we have:

\bea\label{sola}
&&\langle a^{\dagger}_{\vec{p} - \vec{q}} a_{\vec{p}}\rangle_t = -i \sum_{\vec{k}}
C_{\vec{k}} \int_{-\infty}^{t} d t' \langle (a^{\dagger}_{\vec{p} -
  \vec{q}} a_{\vec{p}- \vec{k}} - a^{\dagger}_{\vec{p} - \vec{q} + \vec{k}}
a_{\vec{p}})(b_{\vec{k}} +
b^{\dagger}_{-\vec{k}})\rangle_{t'} \nn\\
&& ~~~~~~\exp\[-i
(\epsilon_{\vec{p}} - \epsilon_{\vec{p}-\vec{q}})(t - t') -
\frac{i(q^i q^j - 2 p^i q^j) }{2 m} \int_{t'}^{t} d t'' h_{i j}(t'')\]\, .
\eea

\nin
Finally, inserting Eq.(\ref{sola}) in Eq.(\ref{eomb}), the 
time evolution
equation for $\langle b_{\vec{q}}\rangle$ becomes:

\bea\label{eomb1}
&&\frac{\pa}{\pa t} \langle b_{\vec{q}}\rangle_t + i \omega_{\vq}
\langle b_{\vec{q}}\rangle_t
= -i \frac{\alpha_{-\vq} \Omega^2 h(t)}{\sqrt{\omega_{\vq}}}
+\sum_{\vec{p}, \vec{k}} C_{\vec{q}} C_{\vec{k}} \int_{-\infty}^{t}
d t' \nn\\ 
&&\[
\langle a^{\dagger}_{\vec{p} - \vec{q} + \vec{k}} a_{\vec{p}} (b_{\vec{k}} +
b^{\dagger}_{-\vec{k}})\rangle_{t'}\right.
\left.- \langle a^{\dagger}_{\vec{p} - \vec{q}}
a_{\vec{p}-\vec{k}} (b_{\vec{k}} +
b^{\dagger}_{-\vec{k}})\rangle_{t'}\]\times \nn\\
&&\exp\[-i
(\epsilon_{\vec{p}} - \epsilon_{\vec{p}-\vec{q}})(t - t')\right.
\left. - \frac{i}{2
m} (q^i q^j - 2 p^i q^j) \int_{t'}^{t} d t'' h_{i j}(t'')\]\, .
\eea

If we now neglect the electron-phonon interaction in Eq.(\ref{eomb1}), 
i.e. we consider the zero-th order in the $C$'s, 
we have (for the given boundary conditions):
\be\label{sol0b1}
\langle b_{\vec{q}}\rangle^{^{(0)}}_{t} = \frac{\alpha_{-\vec{q}} \Omega^2
  A}{\sqrt{\omega_{\vq}}} \cdot
\frac{\omega_{\vq} \cos(\Omega t) - i
 \Omega \sin(\Omega t)}{\Omega^2 - \omega_{\vq}^2}\, ,
\ee
where $A=A_{zz}$ is the amplitude of the component of the field in the
longitudinal direction. For $\Omega \sim\omega_{\vq}$, this solution has 
the expected resonant form, with a very large amplitude. Strictly
speaking, for $\Omega = \omega_{\vq}$, the
r.h.s. of Eq.({\ref{sol0b1}) is divergent. We note, however, 
that the well known phenomenological Breit--Wigner shape 
is obtained adding an imaginary part to 
$\omega_{\vq}$. As we shall see below, this damping factor, which 
is always added on phenomenological grounds to the equation that 
describes the bar oscillations, from a microscopic point of
view is mainly due to the electron viscosity, i.e. to the 
interaction of the electrons with the phonons. In more technical
terms, it comes from the renormalization of $\omega_{\vq}$ due to the
electron--phonon interaction.

We observe now that for the correlator 
$\langle a^{\dagger}_{\vec{p} - \vec{q} + \vec{k}} a_{\vec{p}}
b_{\vec{k}}\rangle_t$, as well as for the other similar correlators that
appear in the r.h.s. of Eq.(\ref{eomb1}), the assumption that in the
infinite past the electron-phonon interaction is absent clearly
amounts to the condition:

\be\label{splitting1}
\langle a^{\dagger}_{\vec{p} - \vec{q} + \vec{k}} a_{\vec{p}}
b_{\vec{k}}\rangle_{-\infty} = \langle a^{\dagger}_{\vec{p} - \vec{q} + \vec{k}}
a_{\vec{p}}\rangle_{-\infty}\langle b_{\vec{k}}\rangle_{-\infty}\, .
\ee

At the lowest order in the $C$'s, the above factorisation property
is also satisfied by
$\langle a^{\dagger}_{\vec{p} - \vec{q} + \vec{k}} a_{\vec{p}}
b_{\vec{k}}\rangle_t$ at any $t$:

\be\label{splitting}
\langle a^{\dagger}_{\vec{p} - \vec{q} + \vec{k}} a_{\vec{p}}
b_{\vec{k}}\rangle_t = \langle a^{\dagger}_{\vec{p} - \vec{q} + \vec{k}}
a_{\vec{p}}\rangle_t \langle b_{\vec{k}}\rangle_t\, .
\ee

\nin
This approximation (and the other similar ones) is
crucial to find the desired truncation of the above mentioned
infinite system of differential equations.

Inserting Eq.(\ref{splitting}) in Eq.(\ref{eoma}), 
we see that, to the lowest order, the correlators
$\langle a^{\dagger}_{\vec{p}-\vec{q}}a_{\vec{p}}\rangle$ are
time-independent constants. More precisely, they all vanish unless
$\vec{q}=0$, in which case
$\langle a^{\dagger}_{\vec{p}}a_{\vec{p}}\rangle =n_{\vec{p}}$,
where $n_{\vec{p}}$ is the occupation number for the states with
momentum $\vec{p}.$

Therefore, if we limit ourselves to consider in Eq.(\ref{eomb1}) 
only terms up to $O(C^{2}_{\vec{q}})$, we need to
keep only those terms with $\vec{k}=\vec{q}$. Then, inserting
the expression (\ref{hij}) for $h_{ij}$ in Eq.(\ref{eomb1}) we have:

\bea\label{eomb2}
&&\frac{\pa}{\pa t} \langle b_{\vec{q}}\rangle_t + i \omega_{\vq}
\langle b_{\vec{q}}\rangle_t = 
-i \frac{\alpha_{-\vq} \Omega^2
A \cos (\Omega t)}{\sqrt{\omega_{\vq}}}\nn\\
&& +~ C^2_{\vec{q}} \sum_{\vec{p}} (n_{\vec{p}} -
n_{\vec{p} - \vec{q}}) \int_{-\infty}^{t} d t' \(\langle
b_{\vec{q}}\rangle_{t'} + \langle
b^{\dagger}_{-\vec{q}}\rangle_{t'}\)\cdot\nn\\
&&\cdot
\exp\[-i
(\epsilon_{\vec{p}} - \epsilon_{\vec{p}-\vec{q}})(t - t') - \frac{i
  A_{i j}}{2 m} (q^i q^j - 2 p^i q^j) \int_{t'}^{t} d t''
\cos(\Omega t'')\]  .
\eea

It is now convenient to move to Fourier space and write:

\be\label{ftransf}
\langle b_{\vec{q}}\rangle_t = \int_{-\infty}^{\infty} d \omega
B_{\vec{q}}(\omega) e^{-i \omega t} \, .
\ee

\nin
Note that Eq.(\ref{eomb2}) also contains the term
$\langle b^{\dagger}_{-\vec{q}}\rangle_{t}$ and that,  
from Eq.(\ref {ftransf}), we have:

\be\label{ftransf+}
\langle b^{\dagger}_{-\vec{q}}\rangle _t = \int_{-\infty}^{\infty} d \omega
B^{*}_{-\vec{q}}(\omega) e^{i \omega t} \, .
\ee

A relation between $B_{\vec{q}}(\omega)$ and
$B^{*}_{-\vec{q}}(\omega)$ is immediately obtained 
with the help of the
time evolution equation for $\langle
b^{\dagger}_{-\vec{q}}\rangle_{t}$. In fact, taking into account that
the electron-phonon coupling constants are real, and that
$C_{\vec{q}}=C_{-\vec{q}}$, we have:

\be
i \frac{\pa}{\pa t} \langle b^{\dagger}_{-\vec{q}}\rangle_t + \omega_{\vq}
\langle b^{\dagger}_{-\vec{q}}\rangle_t   = -i \frac{\pa}{\pa t} 
\langle b_{\vec{q}}\rangle_t +
\omega_{\vq} \langle b_{\vec{q}}\rangle_t
\ee

\nin
which implies the relation:

\be\label{rela}
B^{*}_{-\vec{q}}(-\omega) = \frac{\omega_{\vq} -
\omega}{\omega_{\vq} + \omega} B_{\vec{q}}(\omega)\, .
\ee

\nin
Inserting now Eqs. (\ref{ftransf}), (\ref{ftransf+}) and (\ref{rela})
in Eq. (\ref{eomb2}), we find:

\bea\label{fury}
&&\int_{-\infty}^{\infty} d \omega e^{- i \omega t} (-i \omega +i
\omega_{\vec{q}}) B_{\vec{q}}(\omega) = C^2_{\vec{q}}
\sum_{\vec{p}} (n_{\vec{p}} -
n_{\vec{p} - \vec{q}})\nn\\
&&\sum_{n, m = -\infty}^{\infty}
J_n (a)  J_m (a)\int_{-\infty}^{\infty} d \omega \int_{-\infty}^{t} d t' e^{- i \omega t'}
\cdot \nn\\
&& e^{-i n \Omega t} e^{i m \Omega t'} e^{-i (\epsilon_{\vec{p}}
- \epsilon_{\vec{p}-\vec{q}})(t - t')}
\cdot\(B_{\vec{q}}(\omega) +
\frac{\omega_{\vec{q}} - \omega}{\omega_{\vec{q}} + \omega}
B_{\vec{q}}(\omega)\)\nn\\
&&-i \frac{\alpha_{-\vq} \Omega^2 A}{2
  \sqrt{\omega_{\vq}}} \int_{-\infty}^{\infty} d \omega
\(\delta(\omega - \Omega) + \delta(\omega + \Omega)\) \, ,
\eea
where $J_n(a)$ are Bessel functions, whose argument $a$ is:

\be\label{argbes}
a=\frac{A_{i j} (q^i q^j - 2 p^i q^j)}{2 m \Omega},
\ee

\nin
and we have used the relation:

\bea
e^{i a \sin \theta} = \sum_{n = -\infty}^{\infty} J_n(a)
e^{i n \theta}\, .\label{bessel}
\eea

The time integral in the r.h.s of Eq.(\ref{fury}) is easily
performed. Isolating the Fourier coefficients, we obtain
($ -i\varepsilon$, as usual, is the converging factor
for the time integral, i.e. it implements the boundary
conditions):

\bea\label{eqB}
&&(\omega - \omega_{\vec{q}}) B_{\vec{q}}(\omega) =  
\frac{\alpha_{-\vq} \Omega^2 A}{2
\sqrt{\omega_{\vq}}}
\(\delta(\omega - \Omega) + \delta(\omega + \Omega)\)~~~~~~~~~~~~\nn\\
&&+C^2_{\vec{q}}
\sum_{n, m = -\infty}^{\infty} \sum_{\vec{p}} J_n (a) J_m (a)
\frac{ n_{\vec{p}} - n_{\vec{p} - \vec{q}}}{- \omega +
n \Omega + \epsilon_{\vec{p}} - \epsilon_{\vec{p}-\vec{q}} - i
\varepsilon}\nn\\
&&\(B_{\vec{q}}(\omega + (m - n)\Omega) +
\frac{\omega_{\vec{q}} - \omega +(n - m)\Omega}{\omega_{\vec{q}} +
\omega + (m - n)\Omega} B_{\vec{q}}(\omega + (m - n)\Omega)\) .
\eea

\nin
From Eq.(\ref{eqB}) we can obtain  
$B_{\vec{q}}(\omega)$ up to $O(C_{\vec{q}}^2)$. To the lowest order,
i.e. by keeping only the first term in the r.h.s. of this equation, we have:

\be\label{eqBzero}
B_{\vec{q}}^{(0)}(\omega) =  
\frac{\alpha_{-\vq} \Omega^2 A}{2
\sqrt{\omega_{\vq}}\,(\omega - \omega_{\vec{q}}) }
\(\delta(\omega - \Omega) + \delta(\omega + \Omega)\)\, ,
\ee

As it can be immediately verified, this is nothing but 
the Fourier transform of Eq.(\ref{sol0b1}). To get the 
$O(C_{\vec{q}}^2)$ correction to Eq.(\ref{eqBzero}), we have 
to insert in Eq.(\ref{eqB}) the series:  
$B_{\vq}(\omega)=B_{\vq}^{(0)}(\omega)+C_{\vq}^2B_{\vq}^{(1)}(\omega)+
O(C_{\vq}^4)$, and keep only terms up to $O(C_{\vq}^2)$. 
Here we are rather interested in the renormalization of the phonon 
frequency $\omega_{_{\vec{q}}}$, let us call $\omega^r_{_{\vec{q}}}$
the renormalised frequency,  due to the electron--phonon interaction. 
In the following we shall see how $\omega^r_{_{\vec{q}}}$ can be 
obtained with the help of Eq.(\ref{eqB}).

We could compute $\omega^r_{_{\vec{q}}}$ directly by 
considering the renormalization of the phonon propagator. As we are interested 
in the $O(C_{\vec{q}}^2)$ correction to $\omega_{_{\vec{q}}}$, 
we only need to keep diagrams with one electron--hole loop.
However, the electrons also interact with the external 
gravitational field (see the first term on the r.h.s. of 
Eq.(\ref{ham-tot})). Therefore, in addition to the usual fermionic 
loop, an infinite series of one--loop diagrams with 
the insertion of one, two, ... gravitational external lines is generated.
These diagrams are all of $O(C_{\vec{q}}^2)$. 

Equivalently, following a standard procedure \cite{eps1}, the same 
result can be obtained from Eq.(\ref{eqB}) once we perform 
the following steps. First of all we note that the 
first term on the r.h.s. of this equation cannot contribute to the 
renormalization of $\omega_{_{\vq}}$. In fact, as it is due to the 
resonant coupling between the external gravitational field and the 
phonons (see the last term on the r.h.s. of 
Eq.(\ref{ham-tot})), this term does not enter in the loop 
correction of the phonon propagator.

If we now: (i) ignore this term in Eq.(\ref{eqB}); (ii) consider 
a value of $\omega$ such that 
$\omega = \omega_{\vec{q}} + O(C_{\vec{q}}^2)$; (iii)  
keep only terms up to $O(C_{\vec{q}}^2)$, we can write Eq.(\ref{eqB}) 
as:

\be\label{reny}
(\omega - \omega_{\vec{q}} - \Pi (\omega)) B_{\vec{q}}(\omega)=0
\ee

\nin
The precise form of $\Pi(\omega)$ is given below 
(see Eqs.(\ref{reno}) and (\ref{renogr})). Apart from harmless 
factors, the coefficient in front  of $B_{\vec{q}}\,(\omega)$ is 
the inverse phonon propagator up to $O(C_{\vec{q}}^2)$. 
Therefore, the one--loop corrected phonon frequency, which is nothing 
but the pole of the phonon propagator, is given by that value of 
$\omega$ for which this coefficient vanishes. 
  
To begin with, let us consider the case when the gravitational 
wave is absent. As in this case $a=0$, all the Bessel functions,
with the exception of $J_0$ ($J_0(0)=1)$, vanish. Following the 
procedure outlined above, neglecting higher order terms in
$C^2_{\vec{q}}$ and dropping the factor $B_{\vec{q}}$, from 
Eq.(\ref{eqB}) we get:

\be\label{reno}
\omega^r_{_{\vq}} = \omega_{_{\vq}} + C^2_{\vec{q}}
\sum_{\vec{p}}
\frac{ n_{\vec{p}} - n_{\vec{p} - \vec{q}}}{- \omega_{_{\vq}}+
\epsilon_{\vec{p}} - \epsilon_{\vec{p}-\vec{q}} - i
\varepsilon}  \, .
\ee

\nin
This is the well known result for the renormalization of
$\omega_{_{\vq}}$ \cite{fetter} in the absence of external fields
(see section \ref{atten} below for the detailed computation), where
we easily recognise the one--loop structure of the second 
term on the r.h.s. of Eq.(\ref{reno}). Note that the negative of 
the imaginary part of $\omega^r_{_{\vq}}$, 
$\Gamma_{\vq} = -{\rm Im}\,\omega^r_{_{\vq}}$,  gives the 
$O(C_{\vec{q}}^2)$ contribution to the damping factor due to the 
electron-phonon interaction. This term, as we have anticipated, 
modifies the delta--like structure of Eq.(\ref{sol0b1}) giving 
to this amplitude the expected Breit--Wigner shape, the maximum being 
attained for $\Omega={\rm Re}\,\omega^r_{_{\vq}}$.

The same steps can be repeated for the case when the
gravitational wave is present. As we only keep terms up to 
$O(C_{\vec{q}}^2)$, the functions $B_{\vec{q}}$ in both members of
Eq.(\ref{eqB}) have to be replaced with their lowest order 
approximations, $B_{\vec{q}}^{(0)}$. Consequently,
the only non vanishing terms in the r.h.s. of  Eq.(\ref{eqB}) are
those for which $n=m$. Moreover, as it was also the case when the 
gravitational wave was absent, the last term of  Eq.(\ref{eqB}),
which is of higher order in $C_{\vec{q}}^2$, has to be neglected. 
Therefore, dropping again $B^{(0)}_{\vec{q}}$ from both sides of 
Eq.(\ref{eqB}), we have: 

\be\label{renogr}
\omega^r_{_{\vq}} = \omega_{\vq} + C^2_{\vec{q}}
\sum_{n = -\infty}^{\infty} \sum_{\vec{p}} \[ J^2_n (a) \frac{n_{\vec{p}}
- n_{\vec{p} - \vec{q}}}{- \omega_{\vq} + n \Omega +
\epsilon_{\vec{p}} - \epsilon_{\vec{p}-\vec{q}} - i \varepsilon}\]\, .
\ee

As we have anticipated, in Eq.(\ref{renogr}) we see that the 
$O(C^2_{\vec{q}})$ correction to $\omega_{\vq}$ contains
an infinite series of one--loop terms due to the interaction of 
the electrons with the gravitational wave. Eq.(\ref{renogr})
is the result we were looking for, namely 
the renormalised $\omega_{_{\vq}}$ ``dressed" by the interaction
with the external gravitational field. This expression clearly 
contains also the contribution to $\omega^r_{_{\vq}}$ which does 
not depend on this interaction, which is nothing but the result 
in Eq.(\ref{reno}). It is easily found in the $n=0$ term of this 
series once we consider the first term of the expansion of $J^2_0(a)$ 
in powers of $a$ (see Eq.(\ref{besse}) below). 

Our aim is to investigate the impact of this additional ``dressing" 
of $\omega^r_{_{\vq}}$ on the oscillations of the gravitational 
antenna. To this end, we compute in the next section the damping factor 
$\Gamma_{\vq} = - {\rm Im}\,\omega^r_{_{\vq}}$. Following the notation
introduced in section \ref{elmw}, we indicate the 
attenuation factor in the absence of the gravitational wave with 
$\Gamma^0_{\vec{q}}$ and the gravitational contribution with  
$\gamma_{\vec{q}}$, so that the total attenuation factor is: 
$\Gamma_{\vec{q}}=\Gamma^0_{\vec{q}} +\gamma_{\vec{q}}$.

\section{The attenuation factor}\label{atten}

This section is devoted to the computation of the attenuation 
(or damping) factor $\Gamma_{\vec{q}}$.  
In order to compare our estimates of physical quantities with 
experimentally known values, in the following we consider the specific 
example of the antenna EXPLORER\cite{gwexp} operating at CERN. Taking the 
imaginary part of Eq.(\ref{renogr}), we have:

\bea\label{damp}
\Gamma_{\vq} &=&- {\rm Im}\,\omega^r_{\vec{q}} =\Gamma^0_{\vec{q}} 
+\gamma_{\vec{q}} \nn\\ 
&=&\pi C^2_{\vec{q}}
\sum_{n = -\infty}^{\infty} \sum_{\vec{p}}  J^2_n
(a) \, (n_{\vec{p} -
\vec{q}} - n_{\vec{p}})\, 
\delta \(\epsilon_{\vec{p}}
- \epsilon_{\vec{p}-\vec{q}} - \omega_{\vq} + n \Omega\) \, .
\eea

Due to the weakness of the gravitational wave, $a$, the dimensionless
argument of the Bessel functions, is a small number. Therefore, 
we can expand these functions in powers of $a$ and limit 
ourselves to consider terms up to the lowest non trivial order in $a$. 
This amounts to keep only the Bessel functions with $n=0, \; \pm 1$. 
In fact, it is:

\bea\label{besse}
J_0^2(a)&=&1-\frac{a^2}{2}+\cdots\;,\nn\\
J^{2}_{\pm 1}(a)&=&\frac{a^{2}}{4}+\cdots\;,
\eea

\nin
where the dots indicate higher power terms in $a$, and 
the expansion of all the other Bessel functions $J^{2}_{n}(a)$ with  
$|n|\geq 2$ starts with higher powers of $a$. We know from 
Eq.(\ref{argbes}) that the quantity $a$ is a function of $\vec p$ : 
$a=a(\vp)$. For reasons that will be immediately clear, in the following
expression it is convenient to indicate explicitly this dependence. 
Keeping in Eq.(\ref{damp}) only terms up to $O(a^2)$, we have:

\bea\label{eps}
&&\Gamma_{\vq} =\pi C^2_{\vec{q}}\Bigg(
\sum_{\vp}n_{\vp}\[\de(\eps_{\vp+\vq}-\eps_{\vp}-\omega_{\vq})
-\de(\eps_{\vp}-\eps_{\vp-\vq}-\omega_{\vq})  \]
+\sum_{\vp}\frac{n_{\vp}}{4}\times~~~\nn\\
&&
\Big(\[\de(\eps_{\vp+\vq}-\eps_{\vp}
-2\omega_{\vq})+\de(\eps_{\vp+\vq}-\eps_{\vp})-2\de(\eps_{\vp+\vq}
-\eps_{\vp}-\omega_{\vq})\]a^2(\vp+\vq)\nn\\
&& 
-\[\de(\eps_{\vp}-\eps_{\vp-\vq}
-2\omega_{\vq})+\de(\eps_{\vp}-\eps_{\vp-\vq})
-2\de(\eps_{\vp}-\eps_{\vp-\vq}-\omega_{\vq}) \]a^2(\vp)\Big)\Bigg)
\eea

The first term in the r.h.s. of Eq.(\ref{eps}) does not depend on $a$.
It is nothing but $\Gamma_{\vq}^0$. Considering for the time being only 
this term, and performing the summation over $\vp$, we get:

\be\label{epsel}
\Gamma_{\vq}^{0}=\frac{8\pi^3 V C_{\vq}^2}{h^4}
\frac{\omega_{\vq} m^2}{q}\, ,
\ee

\nin
where $V$ is the volume of the bar and $m$ is the electron mass.
Moreover, as we have chosen the bar to lie along the $z$ direction,
in the above expression we have $q=q_{z}$.

Let us focus now our attention on the derivation of $\gamma_{\vq}$, i.e.
on the computation of the two remaining terms in the
r.h.s. of  Eq.(\ref{eps}). As previously said,
the bar lies along the z direction, while the gravitational wave
(in the TT gauge) propagates perpendicularly to the bar, along the
$x$ direction. Therefore, the only non vanishing
components of $A_{ij}$ are: $A_{yy}=-A_{zz}$ and $A_{xy}=A_{yx}$, so
that:  

\be\label{asmall}
p^{i}q^{j}A_{ij}= p_y q_z A_{yz} + p_z q_z A_{zz}=
pq(A_{yz}\sin\theta \sin\phi+A_{zz}\cos\theta),
\ee

\nin
where, to write the last term, we have introduced spherical coordinates
in the $\vp$-space (and noted again that $q_z=q$).  

\nin
Inserting Eq.(\ref{asmall}) in Eq.(\ref{eps}), and proceeding as before,
we find: 

\bea\label{gaqu}
\gamma_{\vq} &=& \frac{V}{h^3}\frac{\pi C^2_{\vec{q}}}
{8 \hbar m^2 (\hbar\omega_{\vq})^2}\int_{0}^{p_{F}}p^2 dp\int_{-1}^{+1}
d(\cos\theta)\int_{0}^{2\pi}d\phi\nn\\
&&\Big(\[(\hbar q)^4 A_{zz}^2+4p^2 (\hbar q)^2(A_{yz}^2sin^2\theta
\sin^2\phi+A_{zz}^2 \cos^2\theta)\] \times\nn\\ 
&&~~\[\de_1+\de_2-\de_3-\de_4-2\de_5+2\de_6\] + \nn\\
&&~~4p(\hbar q)^3 A_{zz}^2
\cos\theta\[\de_1+\de_2+\de_3+\de_4-2\de_5-2\de_6\]\Big)\, ,
\eea

\nin
where we have introduced the compact notations:
\bea\label{compa}
\de_1&=&\delta(\frac{p\hbar q\cos\theta}{m}+\frac{(\hbar q)^2}{2m}-2\hbar
\omega_{\vq}) ~~~\de_2 =\delta(\frac{p\hbar q\cos\theta}{m}+
\frac{(\hbar q)^2}{2m})~~~~~~\nn\\
\de_3&=&\delta(\frac{p\hbar q\cos\theta}{m}-\frac{(\hbar q)^2}{2m}-2\hbar
\omega_{\vq}) ~~~\de_4 =\delta(\frac{p\hbar q\cos\theta}{m}-
\frac{(\hbar q)^2}{2m})\nn\\
\de_5&=&\delta(\frac{p\hbar q\cos\theta}{m}+\frac{(\hbar q)^2}{2m}-\hbar
\omega_{\vq}) \nn\\
\de_6 &=&\delta(\frac{p\hbar q\cos\theta}{m}-
\frac{(\hbar q)^2}{2m}-\hbar\omega_{\vq}) \, .
\eea

\nin
For a typical gravitational antenna we are under the 
conditions\footnote{for instance for EXPLORER 
it is:~~$\hbar\omega_{\vq}= 6\cdot 10^{-31}J$,~~ $p_F=1,22
\cdot 10^{-24}Kg\;m/s$,

$\hbar q=1,1\cdot 10^{-34}Kg\;m/s$.}:

\be\label{conditio}
\frac{(\hbar q)^2}{2m}<l\hbar\omega_{\vq}-
\frac{(\hbar q)^2}{2m}<l\hbar\omega_{\vq}+
\frac{(\hbar q)^2}{2m}<\frac{p_F\hbar q}{m}\qquad (l=1,2) \, ,
\ee

\nin
and the integration of Eq.(\ref{gaqu}) gives:

\bea\label{upsGW}
\gamma_{\vq}=\frac{6\pi^3 V C_{\vq}^2}{h^4}\frac{m^2\omega_{\vq}}{ q}
\[2A_{zz}^2-A_{yz}^2\]\, .
\eea

Finally, since the incoming gravitational wave is not expected to be polarized,
now we have to perform the average over the polarizations. If we call $\beta$
the polarization angle, we can write:

\be
A_{zz}=e_{+}\cos2\beta-e_{\times}\sin2\beta,\qquad
A_{yz}=e_{+}\sin2\beta+e_{\times}\cos2\beta
\ee

\nin
Averaging over $\beta$, we arrive to the result:

\be\label{dump}
\gamma_{\vq}=
\frac{3\pi^3 V C_{\vq}^2}{h^4}\frac{m^2\omega_{\vq}}{q}
(e_{+}^2+e_{\times}^2) \, .
\ee

It is interesting to compare Eq.(\ref{dump}) with the corresponding
damping factor for the electromagnetic case (see Eq.(\ref{gamma}) of
section \ref{elmw}). We note that, differently from this case, the
gravitational contribution to $\Gamma_{\vq}$ is always positive.
The gravitational correction to $\Gamma_{\vq}^0$ can never produce
an amplification of the bar oscillation.

Irrespectively of its sign, however, it is more important to compare
the magnitude of $\gamma_{\vq}$, Eq.(\ref{dump}), with $\Gamma_{\vq}^0$, 
Eq.(\ref{epsel}). Taking for $e_{+}\cong e_{\times}= e$ 
the realistic value of $e \cong 10^{-21}$, we obtain:

\be
\frac{\gamma_{\vq}}{\Gamma^0_{\vq}}\cong 10^{-42}\, .
\ee

\nin
Clearly, the gravitational correction to $\Gamma_{\vq}^0$ is too 
small to give any appreciable contribution to $\Gamma_{\vq}$. 

Few comments are in order. First of all we note that such a result, 
which could be expected on the basis of the extreme weakness of the 
gravitational wave amplitude, entirely justifies the common  
assumption according to which the only physically relevant mechanism 
for the transfer of energy between the incoming 
wave and the gravitational antenna is the direct resonant coupling 
between the wave and the bar normal modes. 

We also note that, as the loop integrals extend up to the Fermi momentum, 
we could expect that correction to $\Gamma^0_{\vq}$ depends on $p_{_F}$. 
Performing the momentum integrals in Eq.(\ref{gaqu}), however, we see that
the terms that individually depend on $p_{_F}$ cancel each others. 

One could a priori think that, would these cancellations  
not occur, the correction to $\Gamma^0_{\vq}$ could turn 
out more significative. 
However, the value of the Fermi momentum (see the footnote on page 20) 
is not sufficiently high to compensate for the extremely low value 
of the gravitational amplitude. In fact, as we can easily see by simple 
inspection of Eq.(\ref{compa}), there are three typical terms 
that come out from the
integration of Eq.(\ref{gaqu}) before the cancellations. They are: 
$\(\frac{p_{_{F}} \hbar q}{m}\)^2$, 
$\(\hbar \omega_{\vq}\)^2$ and 
$\(\frac{\hbar^2 q^2}{2 m}\)^2$. 
While the third term is negligible w.r.t. 
the second one, the first term is six orders of magnitude greater than 
the second. Therefore, the order of magnitude of the
``gravitational damping" $\gamma_{\vq}$, even in the absence of these
cancellations, would only change of a factor $10^6$, and the resulting
value of $\gamma_{\vq}$ ($\sim 10^{-36} s^{-1}$) would  still  be
far too small if compared with $\Gamma^0_{\vq}$.

\section{Summary and Conclusions}\label{conclu}

In the present work, after deriving the  microscopic Hamiltonian for 
the interaction between a gravitational wave and a metallic bar 
(the gravitational antenna), we have studied the contribution to the 
oscillations of the antenna due to the interaction of its electrons 
with the incoming wave. 

More precisely, we have considered the contribution to the damping 
factor $\Gamma_{\vq}$ (the negative of the imaginary part of the 
phonon frequency) 
which comes from the interaction between the electrons of the bar and the 
gravitational external field. It turns out that this term is several 
orders of magnitude smaller than the ordinary factor which is due to 
the electron viscosity (i.e. to the electron-phonon interaction).

As is well known, the gravitational wave interacts directly with the 
normal modes of the bar. This resonant interaction is due to the 
modification of the geodesic distance between the ions of the lattice 
induced by the presence of the gravitational field. The contribution 
$\gamma_{\vq}$ to the damping factor considered in this paper is the 
result of an additional ``indirect" interaction between the 
gravitational wave and the phonons. This additional coupling, which is
due to the electron-graviton interaction, is induced by electron-hole 
loops.  

However, as the incoming wave is extremely weak and the Fermi momentum 
(which is the highest value of momentum running in the loops) is not 
sufficiently high to compensate for this weakness, it should not come 
as a surprise that $\gamma_{\vq}$ turns out to be a negligibly small 
contribution to $\Gamma_{\vq}$.

Finally, it is worth to mention that our findings, while confirming 
the assumption that the only relevant mechanism of interaction between 
a gravitational wave and a metallic bar is the direct resonant 
coupling between the wave and its normal modes, strongly disagree 
with the recent claim that the interaction of the gravitational wave 
with the electrons of the bar enhances the resonant 
cross-section of several orders of magnitude, actually of four orders 
of magnitude\cite{sri}.   

\vskip 30 pt

{\large {\bf Aknowledgments}}
\vskip 10 pt
We would like to thank M. Maggiore, E. Picasso and G. Veneziano for 
many helpful discussions.

\vfill
\eject

\end{document}